\documentclass[published]{JHEP3} 

\JHEP{00(2007)000}

\usepackage{epsfig,multicol,bbm}

\newcommand\fverb{\setbox\pippobox=\hbox\bgroup\verb}
\newcommand\fverbdo{\egroup\medskip\noindent%
            \fbox{\unhbox\pippobox}\ }
\newcommand\fverbit{\egroup\item[\fbox{\unhbox\pippobox}]}
\newbox\pippobox

\newcommand\kzero{{\rm K}_0}
\newcommand\kone{{\rm K}_1}

\def\lsim{\raise0.3ex\hbox{$<$\kern-0.75em\raise-1.1ex\hbox{$\sim$}}}
\def\gsim{\raise0.3ex\hbox{$>$\kern-0.75em\raise-1.1ex\hbox{$\sim$}}}

\newcommand{\rr}{\mbox{\boldmath $r$}}
\newcommand{\rrn}{\mbox{$r$}}

\def\lsim{\raise0.3ex\hbox{$<$\kern-0.75em\raise-1.1ex\hbox{$\sim$}}}
\def\gsim{\raise0.3ex\hbox{$>$\kern-0.75em\raise-1.1ex\hbox{$\sim$}}}

\newcommand{\be}{\begin{equation}}
\newcommand{\ee}{\end{equation}}

\def\beq{\begin{equation}}
\def\eeq{\end{equation}}
\def\beqa{\begin{eqnarray}}
\def\eeqa{\end{eqnarray}}

\newcommand{\rb}{\mbox{\boldmath $b$}}

\def\gappeq{\mathrel{\rlap {\raise.5ex\hbox{$>$}}
{\lower.5ex\hbox{$\sim$}}}}

\def\lappeq{\mathrel{\rlap{\raise.5ex\hbox{$<$}}
{\lower.5ex\hbox{$\sim$}}}}

\def\Toprel#1\over#2{\mathrel{\mathop{#2}\limits^{#1}}}

\title{Saturation Physics  in Ultra High Energy Cosmic Rays: Heavy Quark Production}
\author{V.P. Gon\c{c}alves $^{1,a}$
 and M.V.T. Machado $^{2,b}$ \\
$^1$ Instituto de F\'{\i}sica e Matem\'atica,
Universidade Federal de
Pelotas\\
Caixa Postal 354, CEP 96010-900, Pelotas, RS, Brazil\\
$^2$ Centro de Ci\^encias Exatas e Tecnol\'ogicas, Universidade Federal do Pampa \\
Campus de Bag\'e, Rua Carlos Barbosa. CEP 96400-970. Bag\'e, RS, Brazil\\
    E-mail:  \email{$^a$barros@ufpel.edu.br}, \email{$^b$mmachado.unipampa@ufpel.edu.br},}
\received{}        
\revised{}
\accepted{}        

\abstract{ In this work we estimate the heavy quark production in
the interaction of ultra high energy cosmic rays in the atmosphere,
considering that the primary cosmic ray is a proton or a photon.  At
these energies  the saturation momentum $Q_{\mathrm{sat}}^2(x)$
stays above the hard scale $\mu_c^2=4m_c^2$, implying charm
production probing the saturation regime. In particular, we show
that the  $ep$ HERA data
   presents a scaling   on
$\tau_c \equiv (Q^2+\mu_c^2)/Q_{\mathrm{sat}}^2(x)$. We derive our
results considering the dipole picture and the  Color Glass
Condensate formalism,  which one shows to be  able to describe the
heavy quark production in $\gamma p$ and $pp$
  collisions. Nuclear effects are considered for the scattering of primaries with the air nuclei and we provide a parametrization for the charm and bottom differential cross sections, $d\sigma/dx_F$, which can be used as an input for numerical implementations for lepton flux.  Implications on the flux   of  prompt leptons at the Earth are analyzed and a large suppression is predicted.
}


\begin{document}


\section{Introduction}
Ultra high energy cosmic rays (UHECRs) remain a puzzle in physics.
Although the existence of UHECRs with energies above 10$^{20}$ eV is
now a well-established fact, the theoretical understanding of its
origin and propagation is a subject of  strong interest and intense
discussion \cite{review_cosmic}. In general terms the current
theories for the origin of UHECRs can be classified in two distinct
scenarios. In the Bottom-Up scenario, charged particles are
accelerated from lower energies to high energies in astrophysical
environments. On the other hand, in the Top-Down scenario, the
energetic particles arise from decay of massive particles
originating from physical processes in the early Universe. Another
open question is the  basic composition of the UHECRs.  While in the
Bottom-Up models the primaries are accelerated protons (or nuclei),
 the Top-Down models  predict an increasing photon component at energies
above $10^{19.7}$ eV. Indeed, even in bottom-up models ultra high
energy photons are expected from the GZK process during the cosmic
ray propagation \cite{photonsgzk}. Finally, the interaction of the
UHECRs with the atmosphere nuclei probes the theory of the strong
interactions in a new kinematical range characterized by a center of
mass energy of approximately 500 TeV, which is more than one order
of magnitude larger than the future Large Hadron Collider at
CERN.


An important subject in cosmic ray physics is the flux of prompt
leptons at the Earth which reflects the primary interactions at
energies that can by far exceed the highest available energies. Its
quantification is essential for the cosmic ray physics   as well as
for neutrinos physics. On one hand, the flux of cosmic ray muons in
the atmosphere, underground and underwater provides a way to testing
the inputs of nuclear cascade models, that is, parameters of the
primary cosmic ray flux (energy spectrum, chemical composition, ...)
and particle interactions at high energies \cite{bugaev}. On the
other hand,  the flux of atmospheric neutrinos and muons at very
high energies provides the main background of searches for the muons
neutrinos from extra-galactic neutrino sources in the neutrinos
experiments (e.g. AMANDA, Antares, Nestor), limiting the sensitivity
of any neutrino telescope to astrophysical signals
\cite{review_halzen}.

The flux of prompt leptons is directly associated to the charmed
particle production and its decays, whose estimation is strongly
dependent on the model used to calculate the charm production cross
section and energy spectra
\cite{leptonflux_tig,leptonflux_prs,leptonflux_ggv,leptonflux_mrs}.
It is related to
 the need of extrapolating charm production
data obtained at accelerators energies to the order of magnitudes
higher energies of the relevant cosmic rays collisions. Another
uncertainty present in the estimate of prompt lepton fluxes is
associated to the fact that different authors do not use the same
atmospheric particle showering routines, turning the comparison
among their predictions even more difficult (See e.g. Ref.
\cite{costa}).

Over the past few years,  the heavy quark production at colliders
has received considerable attention. In particular, the
compatibility among perturbative QCD (pQCD) and experimental data
was verified and today one can conclude that even for charm, whose
low mass and larger non-perturbative hadronisation corrections might
cast doubts on the predictive power of the theory, pQCD actually
manages to deliver good results. This framework was used in the
 prompt lepton fluxes calculations presented in Refs. \cite{leptonflux_tig,leptonflux_prs,leptonflux_ggv,leptonflux_mrs}. However, since
the lepton fluxes are strongly dependent on the behavior of the
gluon distribution at small-$x$ \cite{ggv_gluon} and it is
determined by the QCD dynamics, we have that any new dynamical
effect will modify the estimates of the lepton fluxes. Recent
results at the DESY electron-proton collider HERA indicate the
presence of the parton saturation effects (See e.g.
 Ref. \cite{hdqcd}), which modify the linear DGLAP dynamics
\cite{dglap}. Furthermore, a current open question is related to the
possibility of the breakdown of the collinear factorization at
higher  energies due to saturation effects which are expected to be
present in this regime \cite{gelis,nik_nonlinear}. As in the
previous calculations
\cite{leptonflux_tig,leptonflux_prs,leptonflux_ggv,leptonflux_mrs}
the authors have assumed the validity of the collinear factorization
and gluon distributions which are solution from the DGLAP evolution
equation, we believe that a new study of the prompt lepton flux is
timely and necessary.

 In this paper we estimate the heavy quark
production considering the present understanding of the high energy
regime  of the theory of strong interactions (For recent reviews see
Ref. \cite{hdqcd}). In this regime, perturbative Quantum
Chromodynamics (pQCD) predicts that the small-$x$ gluons in a hadron
wavefunction should form a Color Glass Condensate (CGC),
 which  is characterized by the limitation on the maximum
phase-space parton density that can be reached in the hadron
wavefunction (parton saturation), with the transition being
specified  by a typical scale, which is energy dependent and is
called saturation scale $Q_{\mathrm{sat}}$. In order to estimate the
saturation effects we calculate the heavy quark production using the
color dipole approach, which gives a simple unified picture for this
process in photon-hadron and hadron-hadron interactions. Beside charm production we also estimate the  bottom production, since
$B$-hadrons can also contribute to the $\tau$ and $\nu_{\tau}$
fluxes. It is important to emphasize that differently of the charm
quark production in the current accelerators, where the saturation
scale $Q_{\mathrm{sat}}$ is smaller than the typical hard scale,
$\mu_Q = 2 m_Q$, at the energies of interest in this paper, $E >
10^{18}$ eV, we will probe
 for the first time the kinematical regime where $Q_{\mathrm{sat}} >
 \mu_c$. Therefore, at these energies  one can expect a large
 modification of the charm quark total cross sections and,
 consequently, on the flux   of  prompt leptons at the Earth.

Here we restrict ourselves to  estimate the heavy quark production
considering the present understanding of the high energy regime  of
the theory of strong interactions and to predict the magnitude of
the saturation effects in the main quantities which are used as
input in the atmospheric particle shower routines. As the
theoretical predictions of the prompt leptons depend strongly on the
behavior of the total cross section for heavy quark production at
high energies, we believe that this partial calculation allow us to
obtain a reasonable estimate of the magnitude of the saturation
effects on prompt lepton flux at the Earth. Of course,  more
detailed studies are necessary in  future in order to get
precise predictions. This paper is organized as follows. In the next
section we present a brief review of the heavy quark production in
the color dipole picture, demonstrating the direct relation between
the total heavy quark cross sections and the dipole-target cross
section $\sigma_{dip}$. The QCD dynamics is discussed in Section
\ref{section3} and the phenomenological model for $\sigma_{dip}$
used in our calculations is presented. In the Section \ref{section4}
we present our results and our main conclusions are summarized in
the Section \ref{section5}.

\section{Heavy quark production in the dipole picture.}
\label{section2}

 Let us start the analyzes considering the heavy quark
production in photon-hadron interactions at high energies. It  is
usually described in the infinite momentum frame  of the hadron in
terms of the scattering of the photon off a sea quark, which is
typically emitted  by the small-$x$ gluons in the proton. However,
in order to disentangle the small-$x$ dynamics of the hadron
wavefunction, it is more adequate to consider the photon-hadron
scattering in the dipole frame, in which most of the energy is
carried by the hadron, while the  photon has just enough energy to
dissociate into a quark-antiquark pair before the scattering. In
this representation the probing projectile fluctuates into a
quark-antiquark pair (a dipole) with transverse separation $\rr$
long after the interaction, which then scatters off the proton
\cite{nik}. In this approach  the heavy quark photoproduction cross
section  reads as,
\begin{eqnarray}
\sigma\,(\gamma p \rightarrow Q\overline{Q}X)= \int dz\, d^2\rr
\,|\Psi^{\gamma}_T (z,\,\rr,\,Q^2=0)|^ 2 \,\sigma_{dip}(x,\rr) ,
\label{dipapprox}
\end{eqnarray}
where $\Psi^{\gamma}_T(z,\,\rr,\,Q^2)$  is the transverse light-cone
wavefunction of the photon which is given by \cite{nik}
\begin{eqnarray}
 |\Psi_{T}^{\gamma}\,(z,\,\rr,\,Q^2=0)|^2 & = &  \frac{6\alpha_{\mathrm{em}}}{4\,\pi^2} \,
  e_Q^2 \, \left\{[z^2 + (1-z)^2]\, \varepsilon^2 \,K_1^2(\varepsilon \,r)
 +\,  m_Q^2 \, \,K_0^2(\varepsilon\,r)
 \right\}\label{wtrans}
 \end{eqnarray}
where the quantity $\varepsilon^2=z(1-z)\,Q^2 + m^2_Q$
depends on the heavy quark mass, $m_Q$. The $K_{0,1}$ are the McDonald
function.  The variable $\rr$ defines the relative transverse
separation of the pair (dipole), whereas $z$ and $\bar{z}\equiv (1-z)$ is
the longitudinal momentum fractions of the quark and antiquark,
respectively. The dipole cross section, $\sigma_{dip}$,
parameterizes the cross section for the dipole-target
(nucleon or nucleus) interaction. As usual, the Bjorken variable is denoted by
$x$.

Concerning heavy quark hadroproduction, it has been usually described
considering the collinear factorization, with the behavior of the
total cross section at high energies being determined by the
gluon-gluon fusion mechanism  and the gluon distribution at
small-$x$. However, at large energies this process  can be described
in terms of the color dipole cross section, similarly to
photon-hadron interactions, using the color dipole formalism
\cite{npz}. In this approach the total heavy quark production cross
section is given by \cite{npz}
\begin{eqnarray}
\sigma(pp\to Q\bar Q X) & = &
2\int_0^{-\ln(\frac{2m_Q}{\sqrt{s}})}dy\,
x_1G\left(x_1,\mu_F^2\right) \times   \sigma(GN\to Q\bar Q X)\,\,,
\label{ccppdip}
\end{eqnarray}
where  $y=\frac{1}{2}\ln(x_1/x_2)$ is the rapidity of the pair,
$\mu_F\sim m_Q$ is the factorization scale, $x_1G(x_1,\mu_F^2)$ is
the projectile gluon density at scale $\mu_F$ and the partonic cross
section $\sigma(GN\to Q\bar Q X)$ is given by \beq\label{eq:all}
\sigma(GN\to Q\bar Q X) =\int dz \, d^2 \rr \left|\Psi_{G\to Q\bar
Q}(z,\rr)\right|^2 \sigma_{q\bar q G}(z,\rr), \nonumber \eeq with
$\Psi_{G\to Q\bar Q}$ being the pQCD calculated distribution
amplitude , which describes the dependence of the $|Q \bar Q
\rangle$ Fock component on transverse separation and fractional
momentum. It is given by
\begin{eqnarray}\nonumber\label{eq:lcwf}
\left|\Psi_{G\to Q\bar
Q}(z,\rr)\right|^2 =\frac{\alpha_s(\mu_R)}{(2\pi)^2}\left\{m_Q^2\kzero^2(m_Q
r)+ \left[z^2+{\bar{z}}^2\right]m_Q^2\kone^2(m_Q r)\right\},
\end{eqnarray}
 where $\alpha_s(\mu_R)$ is the strong
coupling constant, which is probed at a renormalization scale
$\mu_R\sim m_Q$. Moreover, $\sigma_{q\bar qG}$ is the cross section
for scattering a color neutral quark-antiquark-gluon system on the
target and is directly related with the dipole cross section as
follows
 \beq\label{eq:qqG}
\sigma_{q\bar qG}=\frac{9}{8}\left[\sigma_{dip}(x_2,z \rr)
+\sigma_{dip}(x_2,\bar{z}\rr)\right]
-\frac{1}{8}\sigma_{dip}(x_2,\rr) \nonumber. \eeq
  The basic idea on
this approach is that at high energies a gluon $G$ from the
projectile hadron can develop a fluctuation which contains a heavy
quark pair ($Q\bar Q$). Interaction with the color field of the
target then may release these heavy quarks. In Ref. \cite{rauf} the
equivalence between this approach and the gluon-gluon fusion
mechanism of the conventional
 collinear factorization  has been demonstrated. In particular, the
 dipole predictions are similar to those obtained using  the
 next-to-leading order parton model calculation. Moreover, the
 predictions for heavy quark production agree well with the current
 experimental data.

It is important to emphasize that the color dipole picture is valid
for high energies, where the coherence length $l_c \approx 1/x_2$ is
larger than the target radius. At smaller energies, we should
include, for instance, the quark-antiquark contributions. However,
at high energies the dipoles with fixed transverse separations are
the eigenstates of interaction in QCD, which becomes the
eikonalization an exact procedure, allowing a direct generalization
of the approach to calculate the heavy quark production in
photon-nucleus and proton-nucleus collisions. Moreover, this
approach is the natural framework to include the saturation effects.
Another aspects that should be emphasized are that (i) in this
formalism   a $K$-factor is not necessary, since is includes all
higher-twist   and next-to-leading corrections; (ii) the dipole
cross section is universal, i.e., it is process independent.

\section{High
Energy QCD Dynamics.} \label{section3}

 The  Color Glass Condensate
is
 described by an infinite hierarchy of  coupled evolution equations for the correlators of
Wilson lines \cite{CGC}. In the absence of correlations, the first
equation in the Balitsky-JIMWLK hierarchy decouples and is then
equivalent to the equation derived independently by Kovchegov within
the dipole formalism \cite{KOVCHEGOV}. In the CGC formalism
$\sigma_{dip}$ can be computed in the eikonal approximation,
resulting
\begin{eqnarray}
\sigma_{dip} (x,\rr)=2 \int d^2 \rb \, {\cal{N}}(x,\rr,\rb)\,\,,
\end{eqnarray}
where ${\cal{N}}$ is the  dipole-target forward scattering amplitude
for a given impact parameter $\rb$  which encodes all the
information about the hadronic scattering, and thus about the
non-linear and quantum effects in the hadron wave function. It is
useful to assume that the impact parameter dependence of $\cal{N}$
can be factorized as ${\cal{N}}(x,\rr,\rb) = {\cal{N}}(x,\rr)
S(\rb)$, so that $\sigma_{dip}(x,\rr) = {\sigma_0}
\,{\cal{N}}(x,\rr)$, with $\sigma_0$ being   a free parameter
related to the non-perturbative QCD physics. The Balitsky-JIMWLK
equation describes the energy evolution of the dipole-target
scattering amplitude ${\cal{N}}(x,\rr)$. Although a complete
analytical solution is still lacking, its main properties are known:
(a) for the interaction of a small dipole ($\rr \ll
1/Q_{\mathrm{sat}}$), ${\cal{N}}(\rr) \approx \rr^2$, implying  that
this system is weakly interacting; (b) for a large dipole ($\rr \gg
1/Q_{\mathrm{sat}}$), the system is strongly absorbed and therefore
${\cal{N}}(\rr) \approx 1$. The typical momentum scale,
$Q_{\mathrm{sat}}^2\propto x^{-\lambda}\,(\lambda\approx 0.3)$, is
the so called saturation scale. This property is associated  to the
large density of saturated gluons in the hadron wave function.  In
our analysis we will consider the  phenomenological saturation model
proposed in Ref. \cite{IIM} (including fit with charm quark) which
encodes the main properties of the saturation approaches, with the
dipole cross section  parameterized as follows
\begin{eqnarray}
\sigma_{dip}^{\mathrm{CGC}}\,(x,\rr) =\sigma_0\,\left\{
\begin{array}{ll}
{\mathcal N}_0 \left(\frac{\bar{\tau}^2}{4}\right)^{\gamma_{\mathrm{eff}}\,(x,\,r)}\,, & \mbox{for $\bar{\tau} \le 2$}\,, \nonumber \\
 1 - \exp \left[ -a\,\ln^2\,(b\,\bar{\tau}) \right]\,,  & \mbox{for $\bar{\tau}  > 2$}\,,
\end{array} \right.
\label{CGCfit}
\end{eqnarray}
where $\bar{\tau}=\rr Q_{\mathrm{sat}}(x)$ and the expression for
$\bar{\tau} > 2$  (saturation region)   has the correct functional
form, as obtained  from the CGC formalism \cite{CGC}. For the color
transparency region near saturation border ($\bar{\tau} \le 2$), the
behavior is driven by the effective anomalous dimension
$\gamma_{\mathrm{eff}}\, (x,\,r)= \gamma_{\mathrm{sat}} + \frac{\ln
(2/\bar{\tau})}{\kappa \,\lambda \ln (1/x)}$, where
$\gamma_{\mathrm{sat}}=0.63$ is the LO BFKL anomalous dimension at
saturation limit and $\kappa=9.9$. Hereafter, we label this model
by CGC.



In the case of photon-nucleus and hadron-nucleus interactions we
take the Glauber-Gribov formalism for nuclear shadowing, with an
average mass number $A=14.5$. This is done by replacing
$\sigma_{dip}$ for a nucleon by a nuclear dipole cross section,
\begin{eqnarray} \sigma_{dip}^{nucl}=2\int d^2b\,
\,\left[1-\exp\,\left(-\frac{1}{2}\,AT_A(b)\,\sigma_{dip}\right)\right] \,\,,\label{signuc}
\end{eqnarray}
 where $T_A(b)$ is the nuclear profile function. This approach
leads a reasonable description of the current experimental data for
the nuclear structure function \cite{armesto}.

As a short comment, the present work is based extensively on the Color Glass 
Condensate. However, there are complementary approaches in literature. For instance, in Ref. \cite{Shoshi:2002in} a loop-loop correlation model is able to compute hadronic cross sections in good agreement with cosmic ray data from Fly's eye and Akeno. Moreover, saturation in the impact parameter space can be explicitly investigated in that model.

\FIGURE{\epsfig{file=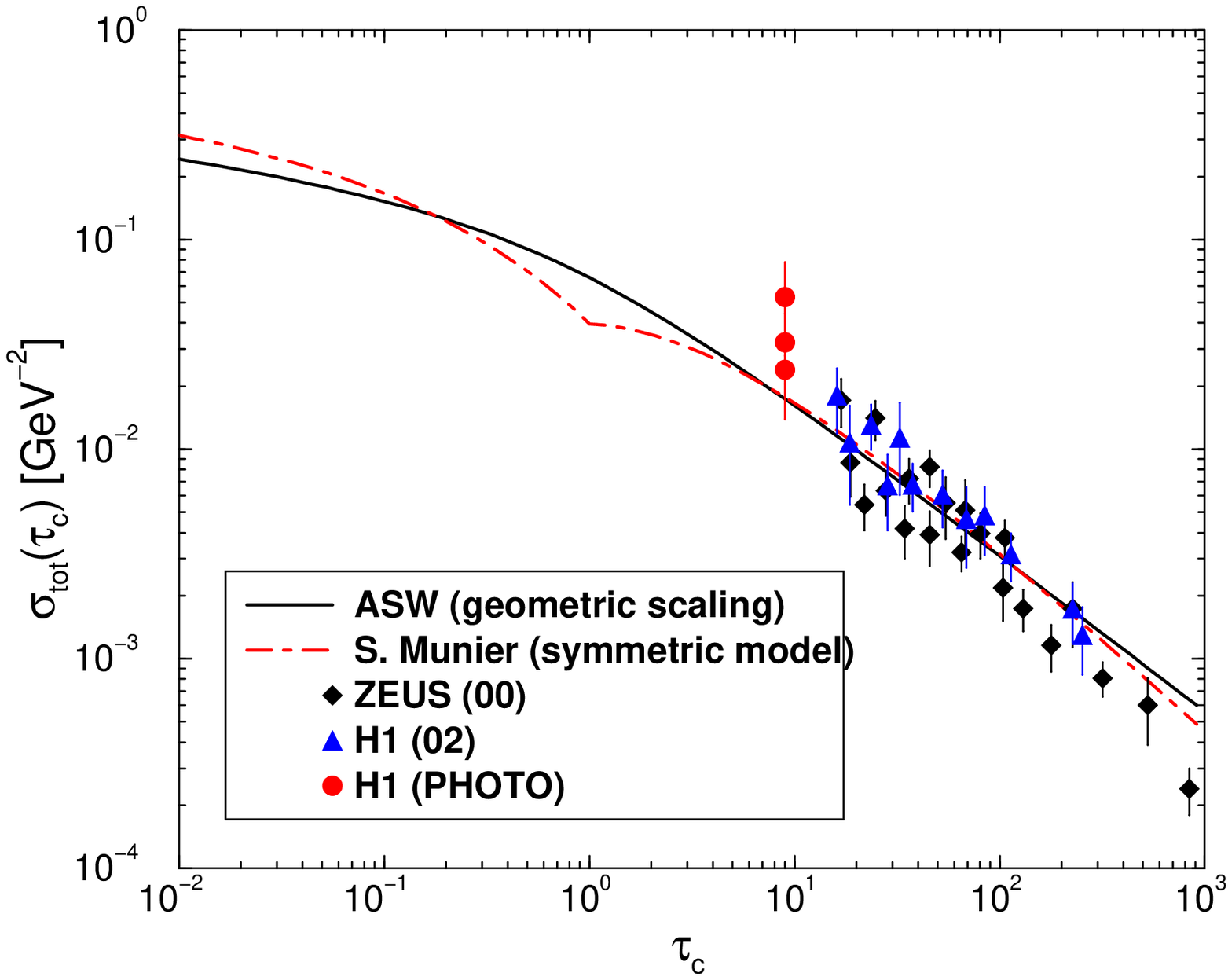,width=10cm}
        \caption[Experimental data on inclusive charm production at
DESY-HERA plotted versus the scaling variable
$\tau_c$.]{Experimental data on inclusive charm production at
DESY-HERA plotted versus the scaling variable $\tau_c$.}%
    \label{fig:1}}

\section{Results for Cosmic Rays}
\label{section4}

 Before presenting our results for the heavy quark
production at very high energies let us consider a basic property of
the saturation physics: the geometric scaling. When applied for the
deep inelastic scattering it means that  the total $\gamma^* p$
cross section at large energies is not a function of the two
independent variables $x$ and $Q$, but is rather a function of the
single variable $\tau = Q^2/Q_{\mathrm{sat}}^2(x)$ as demonstrated
in Ref. \cite{SGK}. In Ref. \cite{prl} we demonstrate that this
property is also present in the charm experimental data at $Q^2 >0$.
In order to include the experimental data for the photoproduction of
charm we have reanalyzed our results and verified that the DESY-HERA
experimental data \cite{datahera} at $Q^2 \ge 0$ presents the
geometry scaling behavior on the variable $\tau_c \equiv
(Q^2+4m_c^2)/Q_{\mathrm{sat}}^2$. It is demonstrated in the Fig.
\ref{fig:1}, where we also present the predictions of two scaling
models \cite{MUNIER,ASW}, which we generalize for charm production.
In the symmetric saturation model \cite{MUNIER}, scaling on $\tau_c$
has been computed before in Ref. \cite{prl} and here it is
generalized for the case where saturation scale is larger than
$\mu_c$. It reads as,
\begin{eqnarray}
 \sigma^{c\bar{c}}_{tot}\,(\tau_c)=\bar{\sigma}\,\left\{
\begin{array}{ll} 1- \exp \left[ -\frac{k}{\tau_c}\left( 1+ \log \,(\tau_c\right)     \right]
, & \mbox{for $\tau_c > 1$}\,, \nonumber \\
\frac{1}{\tau_c}- \frac{1}{\tau_c}\exp \left[ -k\tau_c\left( 1- \log
\,(\tau_c\right)\right],  & \mbox{for $\tau_c  \le 1$}\,,
\end{array} \right.
\end{eqnarray}
The parameters are taken from \cite{MUNIER}, with $k=0.8$, and the
overall normalization gives $\bar{\sigma}=28$ $\mu$b. Furthermore,
we consider the model proposed in Ref. \cite{ASW} (hereafter ASW
model), which relate the $\gamma p (A)$, $p A$ and $AA$ collisions
through the geometric scaling property of the saturation physics.
The parameters for this model has been obtained from a fit to the
small-$x$ $ep$ DESY-HERA data and lepton-hadron data \cite{ASW}.
When generalized for charm production we obtain
\begin{eqnarray}
\sigma^{\gamma^* A}_{c\bar{c}} =  \frac{\pi R_A^2}{\pi R_p^2}\,
\bar{\sigma}_0\,
  \left[ \gamma_E + \Gamma\left(0,\frac{a}{\tau_{A,c}^{b}} \right) +
         \ln \left(\frac{a}{\tau_{A,c}^{b}}\right) \right]\,,
\label{ccphotparam}
\end{eqnarray}
where $\gamma_E$ is the Euler constant, $\Gamma\left(0,\beta\right)$
the incomplete Gamma function and $R_A$ is the nuclear radius. The
nuclear scaling variable for charm production is  $\tau_{A,c} =
\tau_c\left(\frac{R_A^2}{AR_p^2}\right)^{\Delta}$. The parameters
are $a=1.868$, $b=0.746$, $\Delta= 1.266$, $\pi R_p^2= 1.55 $ fm$^2$
and  $\bar{\sigma}_0=20.28$ $\mu$b. Our predictions for charm
production in $\gamma p$ processes are presented in Fig.
\ref{fig:1}.
 The agreement with experimental data is very good, which  allow us
 to propose the  Eq. (\ref{ccphotparam}) as a  theoretical parameterization for the charm production
 in lepton-nuclei interactions at very high energies.


\FIGURE{\epsfig{file=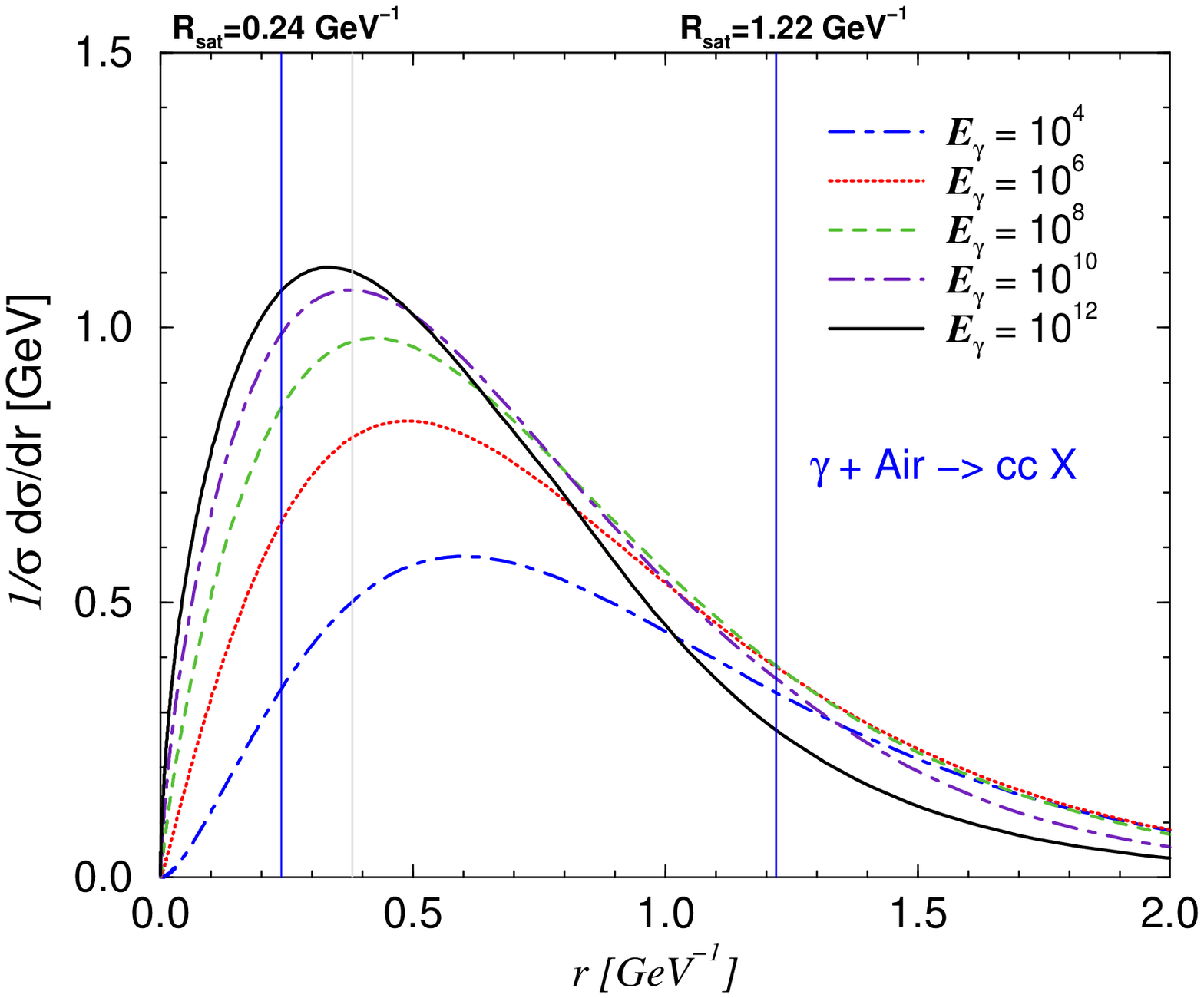, width=9cm}
        \caption[Charm production in $\gamma p$ and $\gamma A$
interactions.]{Profile function for charm production.}%
    \label{fig:2}}


We are ready now to investigate the charm production initiated by primary
cosmic rays photons or protons interacting with atmosphere.
Initially, lets consider the photon-hadron interaction and analyze
the behavior of the profile function, defined by
\begin{eqnarray}
\frac{1}{\sigma_{tot}}\,\frac{d\sigma_{Q\bar{Q}}}{dr} =  \frac{4\pi \rrn}{\sigma_{tot}} \, \int d^2b\,dz\, |\Psi_{T}^{\gamma} (z,\,\rr,Q^2=0)|^2 \,\left[1-\exp\,\left(-\frac{1}{2}AT_A(b)\,\sigma_{dip} (x,\,\rr^2)\right)\right], \label{overlap}
\end{eqnarray}
which allow us to estimate the mean dipole size dominating the
nuclear heavy quark  photoproduction. In Figs. \ref{fig:2} and
\ref{fig:3} are  shown, respectively, the overlap function for the
charm and bottom production as a function of dipole size. They are
computed for the average mass number $A = 14.5$  and for different photon energies. In the
charm case, the distributions are peaked at approximately $r\approx
0.4$ GeV$^{-1}$, whereas for the bottom case this value is shifted
to $r\approx 0.1$ GeV$^{-1}$, which agree with the theoretical
expectation that the $q\overline{q}$ pairs have a typical transverse
size $\approx \, 1/2m_f$ \cite{nik}. Therefore, the main
contribution to the cross section comes from the small dipole sizes,
i. e. from the perturbative regime. In contrast, for light quarks a
broader $r$ distribution is obtained, peaked for large values of the
pair separation, implying that nonperturbative contributions cannot
be disregarded in that case.  We also show in the figures the
corresponding nuclear saturation radius, estimated in a simple way as $R_{\mathrm{sat}}\simeq A^{-1/6}/Q_{\mathrm{sat}}(\bar{x})$, where $\bar{x}= \frac{ 4\,m_Q^2}{W_{\gamma A}^2}$
 and $W_{\gamma A}^2=2 m_N E_\gamma$, for two different values of the photon energy
$E_\gamma$. For the charm case we have that at low energies
($E_\gamma = 10^4$ GeV) the distribution is peaked at values smaller
than the corresponding $R_{\mathrm{sat}}$, which implies that the
saturation effects can be disregarded in this regime. On the other
hand, for high photon energies ($E_\gamma = 10^{12}$ GeV) we have that
the peak occurs at larger values than $R_{\mathrm{sat}}$ and the
saturation effects are expected to contribute for the total cross
section. In contrast, for the bottom case, we have that for all
energy range considered the peaks occur at smaller values than
$R_{\mathrm{sat}}$, which implies that these effects do not
contribute too much for the bottom contribution.

\FIGURE{\epsfig{file=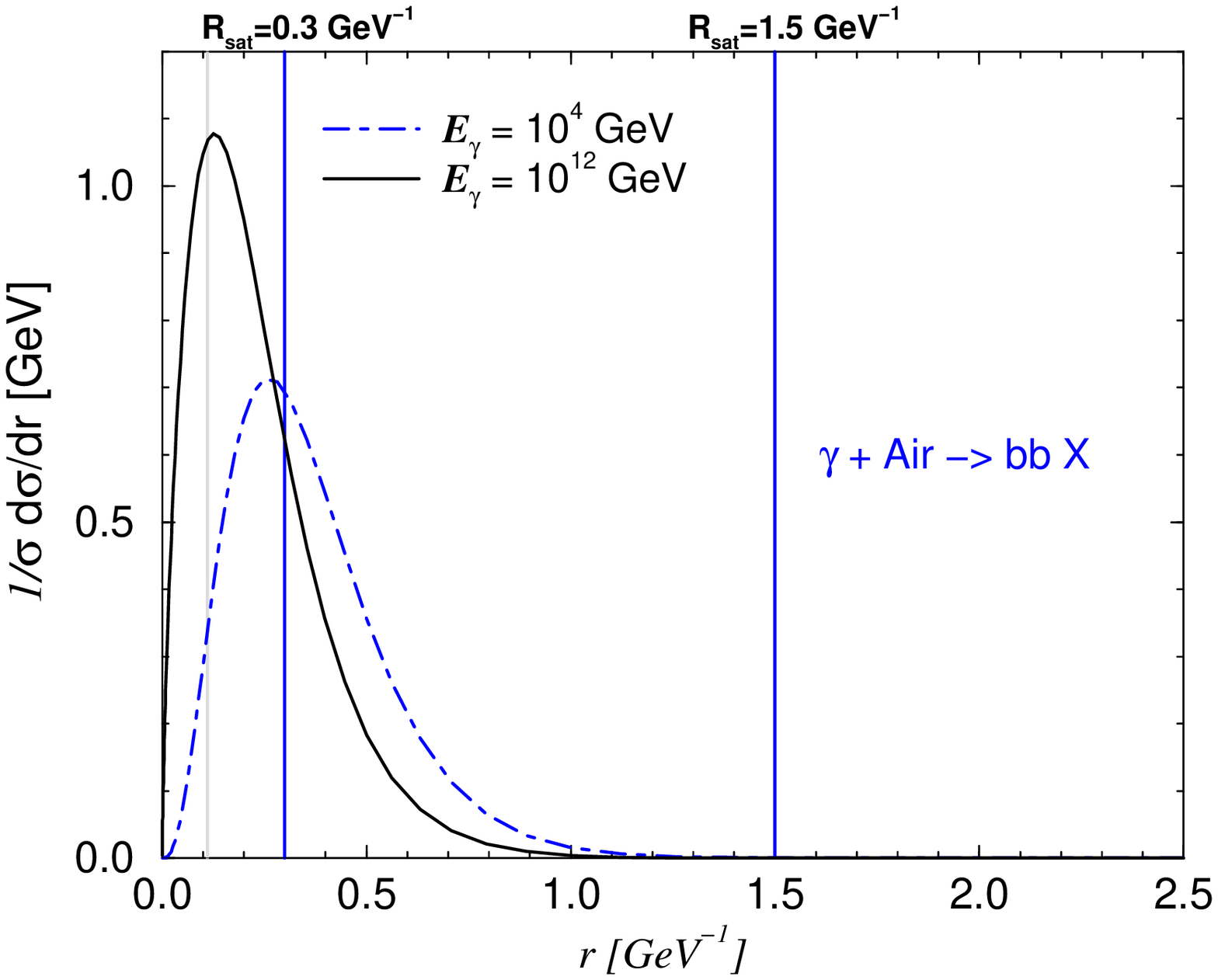, width=9cm}
        \caption[Charm production in $\gamma p$ and $\gamma A$
interactions.]{Profile function for bottom production.}
    \label{fig:3}}

In Figs. \ref{fig:4} and \ref{fig:4b} we show the charm
and bottom  photoproduction cross sections as a function of photon energy, respectively. The
$\gamma p$ interaction is obtained via Eq. (\ref{dipapprox}) with
the dipole-proton cross section given by
$\sigma_{dip}^{\mathrm{CGC}}$. The experimental data from HERA
\cite{datahera} and fixed target collisions \cite{fixedtarget} are
also included for sake of comparison. The theoretical curves roughly describe the data, which have large uncertainties and large errors. We call attention that the CGC model successfully describes the charm and bottom electroproduction data (even at small $Q^2$) and so we believe in its robustness. Therefore, accurate new experimental measurements in this region is of particular importance.

 As explained in the previous section, for the
photon-air interaction we take the Glauber-Gribov formalism for
nuclear shadowing, with an average mass number $A=14.5$ and
$\sigma_{dip}$ replaced by $\sigma_{dip}^{nucl}$. Initially, lets
discuss the nuclear charm photoproduction. In Fig. \ref{fig:4}  the
accelerator data \cite{datahera} and the predictions of our
theoretical parameterization, Eq. (\ref{ccphotparam}),  are also
presented. The energy dependence is mild at $E_{\gamma}\geq 10^5$
GeV , giving $\sigma \propto E_{\gamma}^{\,0.12}$. The
parameterization is numerically equivalent to the complete CGC
dipole calculation, which can be useful as input in Monte Carlo
implementations.  In order to estimate the magnitude of the
saturation corrections to the high energy process, we also show the
result using the color transparency (long dashed curves)
calculation, which corresponds to the leading twist contribution
$\sigma_{dip}\propto r^2\,xG(x,1/r^2)$. The deviation are about one
order of magnitude at $E_{\gamma}\approx 10^{12}$ GeV, increasing
for higher energies. Therefore, in agreement with the previous
analyzes of the profile function, parton saturation plays an
important role in charm production by cosmic rays interactions when
the primary particle is a photon.

\FIGURE{\epsfig{file=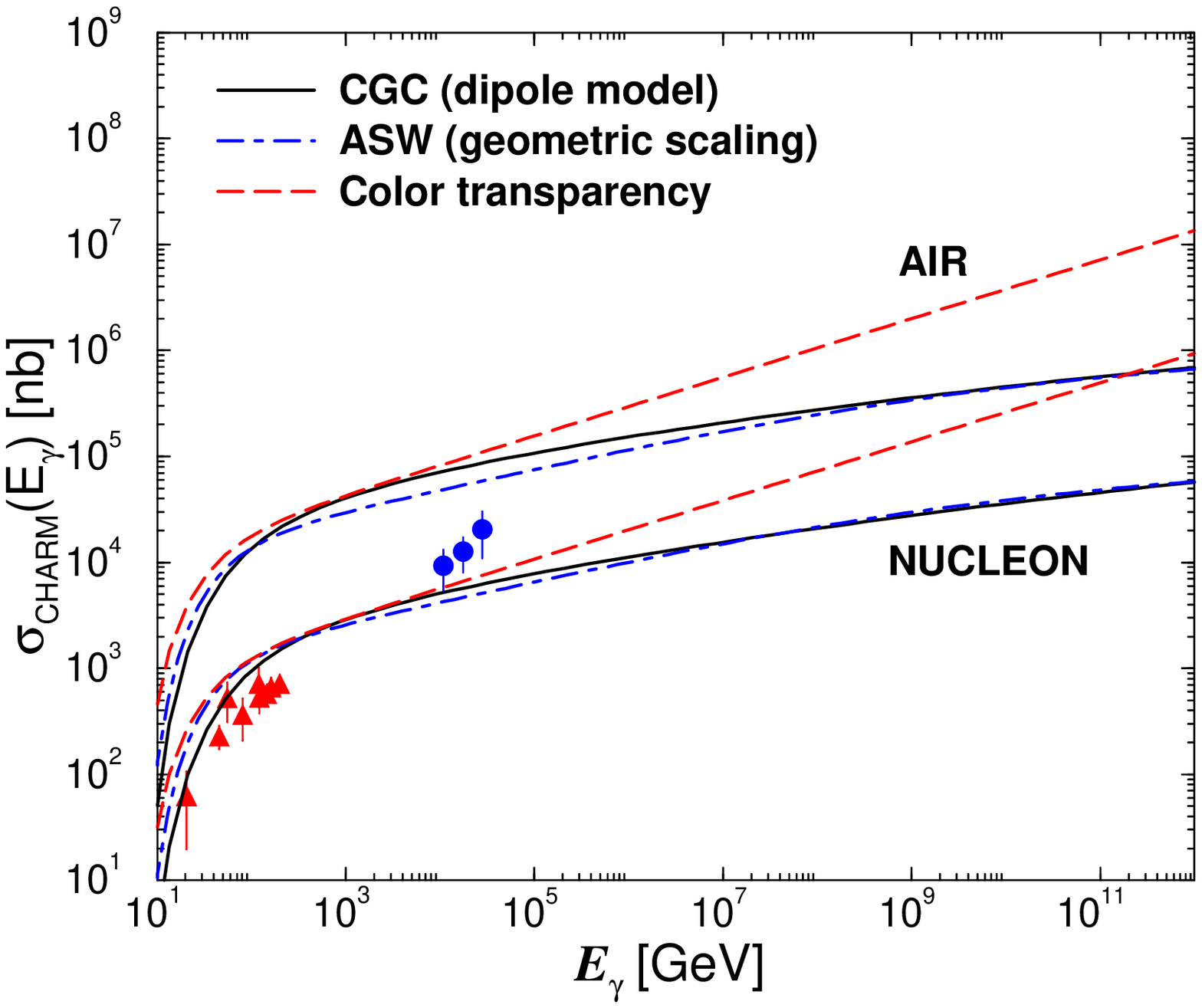,width=10cm}
        \caption[Charm production in $\gamma p$ and $\gamma A$
interactions.]{Charm production in $\gamma p$ and $\gamma A$
interactions. Data from Refs.  \cite{datahera,fixedtarget}.}%
    \label{fig:4}}

Concerning to nuclear bottom photoproduction, our results are
presented in Fig. \ref{fig:4b}. We have that the contribution of the
saturation effects is smaller, being of approximately 10 \% at very
high photon energies ($E_\gamma \approx 10^{12}$ GeV). This result
agree with our previous analyzes of the profile function. An
important aspect is that the  growth with the energy of the bottom
cross section is steeper than the charm case due to distinct
physics which is dominant in the production process (linear versus
saturation physics). It implies that the contribution for the prompt
lepton fluxes  associated to the bottom becomes more important for
larger energies.



\FIGURE{\epsfig{file=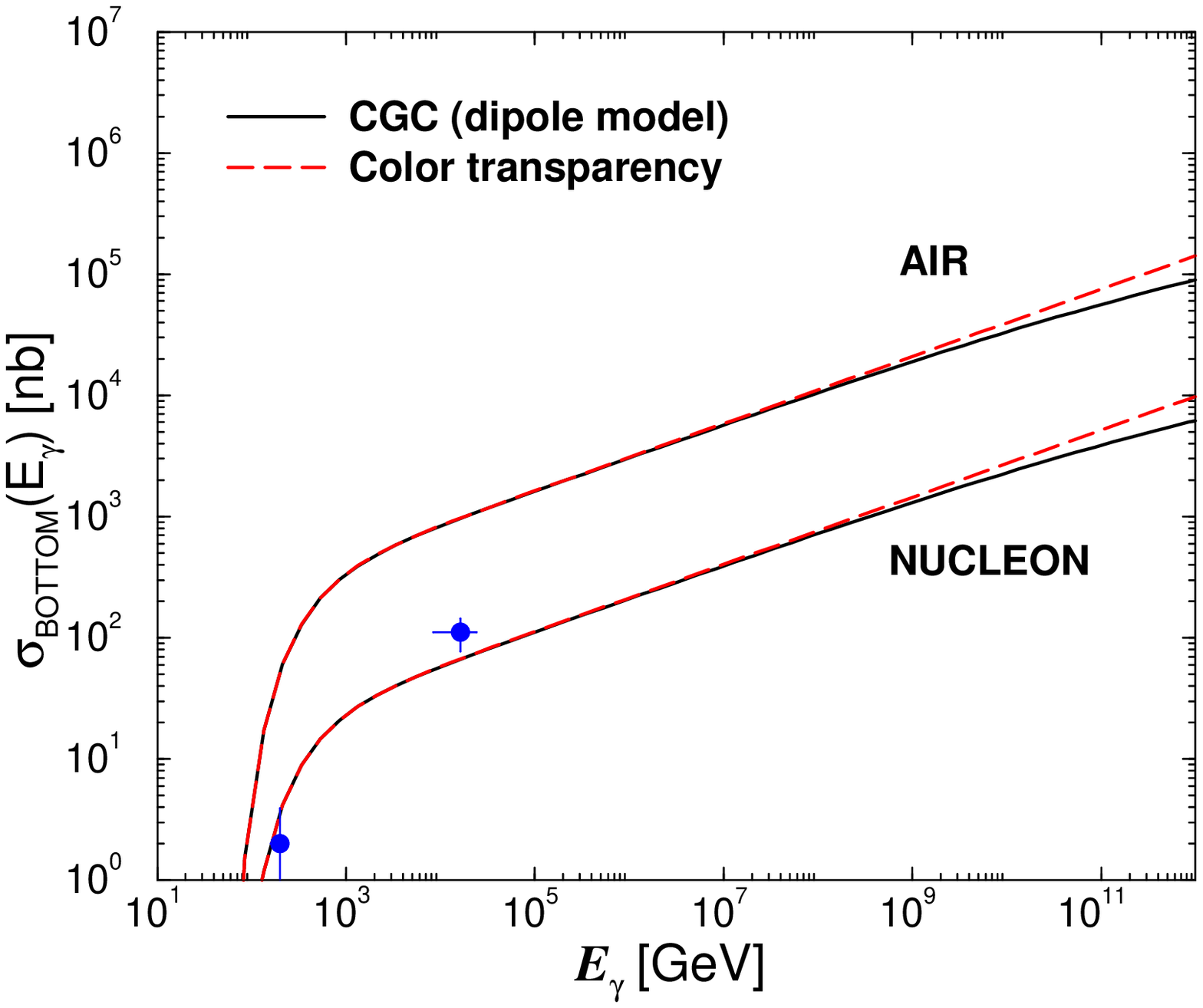,width=10cm}
        \caption[Charm production in $\gamma p$ and $\gamma A$
interactions.]{Bottom production in $\gamma p$ and $\gamma A$
interactions. Data from Refs.  \cite{datahera,fixedtarget}.}%
    \label{fig:4b}}

Lets now consider that the primary cosmic ray is a proton and
estimate the heavy quark production cross section in this case. In
Fig \ref{fig:5} we show the theoretical predictions for charm quark
production in proton-proton and proton-air interactions as a
function of energy of primary cosmic rays. The solid curves
correspond to the calculation using Eq. (\ref{ccppdip}), with gluon
distribution given by the NLO ``physical gluon'' MRST2004
parameterization \cite{mrst2004nlo}. The accelerator data
\cite{accdata} and the color transparency (long dashed curves)
calculation are also presented. We have that the current data at
high energies are well described using the color dipole picture and
that for the energy range of the data the color transparency and CGC
prediction are almost identical. The energy dependence is suddenly
steeper than the photon case, with $\sigma \propto E_{p}^{\,0.17}$
at $E_{p}\geq 10^7$ GeV. The deviation between CGC and leading twist
approaches remains large, being of order of six at $E_{p}\approx
10^{12}$ GeV and increasing with energy. It confirms the importance
of parton saturation corrections in charm quark production via
cosmic rays.

\FIGURE{\epsfig{file=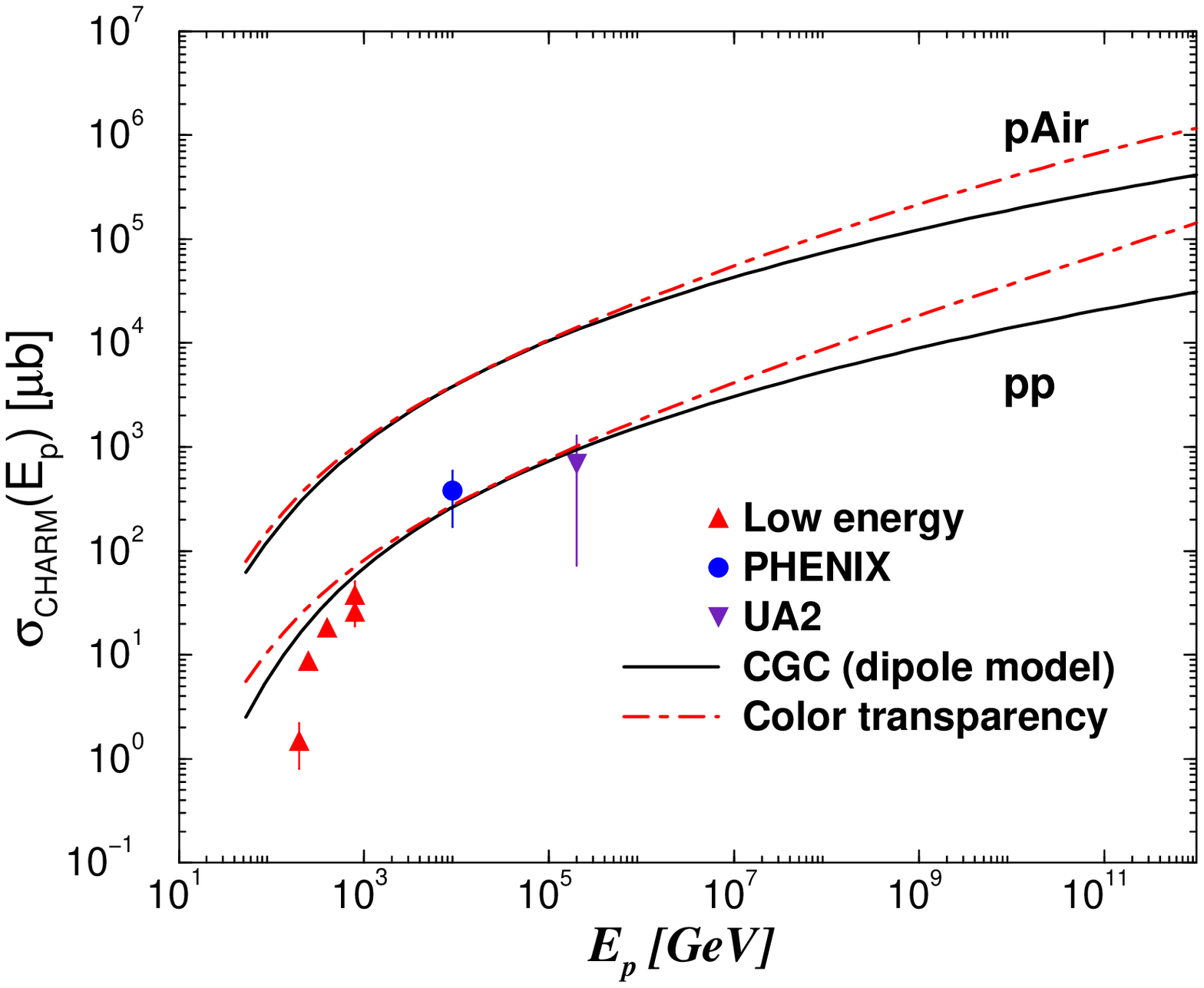,width=10cm}
        \caption[Charm production in $pp$ and $pA$ interactions.]{Charm production in $pp$ and $pA$ interactions. Data from Ref. \cite{accdata}.}%
    \label{fig:5}}

In Ref. \cite{ggv_gluon} the authors have computed the total cross
section for charm production using the collinear factorization,
next-to-leading order corrections and different parton distribution
parameterizations. In particular, the dependence of the predictions
in the behavior of the gluon distribution at small-$x$ was analyzed
in detail. In comparison with those results we have that our
predictions considering the color transparency limit are similar
with the MRST one for $\lambda = 0.2$ (See Fig. 1 in Ref.
\cite{ggv_gluon}). On the other hand, our predictions considering
the saturation effects are very similar to those obtained previously
by Thunman, Ingelman and Gondolo (TIG) \cite{leptonflux_tig}, which
have used an option of PYTHIA by which the gluon distribution is
extrapolated for $x\le10^{-4}$ with $\lambda = 0.08$. As discussed
before, the prompt lepton fluxes are strongly dependent on the charm
cross section. Consequently, we can expect that if our results are
used as input in the atmospheric particle shower routines, the
predicted prompt lepton flux should be similar to the TIG
predictions, which have been considered a lower bound for the flux
of leptons obtained using pQCD. In other words, we have that the
saturation physics on charm quark production implies a suppression
of the prompt lepton fluxes.

\FIGURE{\epsfig{file=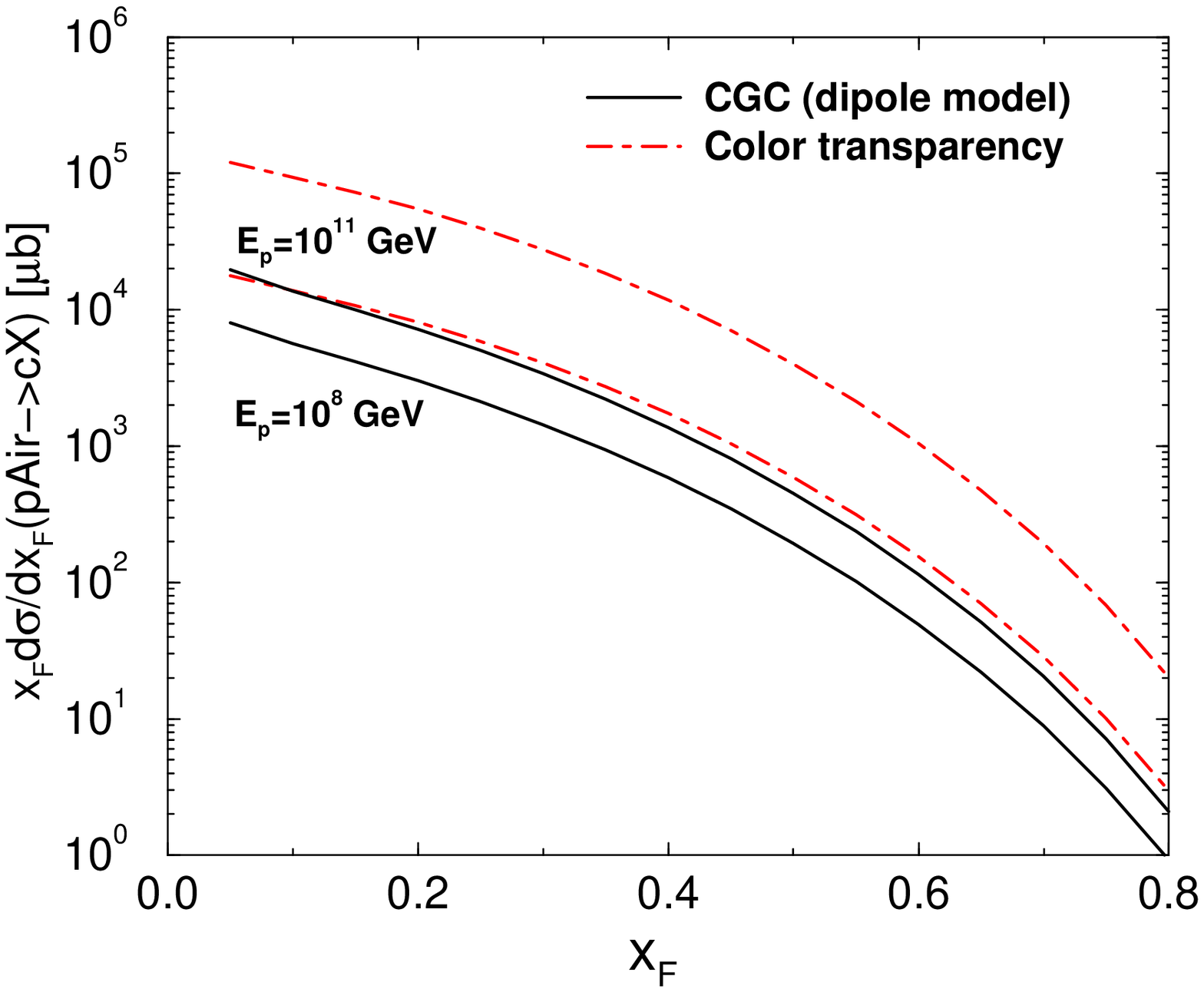,width=10cm}
        \caption[Comparison of the CGC and color transparency predictions for the
        $x_F$-dependence for charm production in  $pA$ interactions.]{Comparison of the CGC and color transparency predictions for the
                $x_F$-dependence for charm production in  $pA$ interactions.}%
    \label{fig:6}}

In what follows we analyze the behavior of  the $x_F$-distribution
for charm production, $x_F d\sigma^c/dx_F$, which is necessary to
calculate the charm production spectrum-weighted moment, also called
production "Z-moment".  The $x_F$-distribution can be directly
obtained from the rapidity distribution in color dipole approach,
which is given by \cite{npz}
\begin{eqnarray}
\frac{d\sigma(pp\to Q\bar Q X)}{dy} = x_1G\left(x_1,\mu_F^2\right)
\sigma(GN\to Q\bar Q X)\,. \label{rapdis}
\end{eqnarray}
Using  that $x_F = x_1 - x_2$ and  $y=\frac{1}{2}\ln(x_1/x_2)$ one
obtain $x_F=\frac{2M_{Q\bar{Q}}}{\sqrt{s}}\,\mathrm{sinh}\,y$, where
$M_{Q\bar{Q}}$ is the invariant mass of the heavy quarks pair and
the momentum fractions are  given by
$x_{1,2}=(\sqrt{x_F^2+\frac{4M_{Q\bar{Q}}}{s}}\pm x_F)/2$.
Consequently, the $x_F$-distribution is
\begin{eqnarray}
\frac{d\sigma(pp\to Q\bar Q
X)}{dx_F}=\frac{1}{\sqrt{\left(2M_{Q\bar{Q}}/\sqrt{s}\right)^2+\left(x_F\right)^2}}\,
\frac{d\sigma(pp\to Q\bar Q X)}{dy} \,.
\end{eqnarray}
 The extension to scattering on nuclei is
straightforward by replacing $\sigma(GN\to Q\bar Q X)$ for
$\sigma(GA\to Q\bar Q X)$. It is important to emphasize that the
heavy quark production actually contains two different phases, which
are not unambiguously separable: the first is the description of the
heavy quark in the hard collision which we have addressed above; the
second is the non-perturbative fragmentation into a heavy hadron,
which cannot be calculated and are in general extracted from the
$e^+ e^-$ data. In our study we do not include the hadronization of
the heavy quarks, since it is model dependent. Of course, in a full
calculation it should be considered.

In Fig. \ref{fig:6} we present this
distribution for two values of the proton energy considering the
saturation effects (CGC) and the color transparency limit. As
expected, the difference between the predictions increases with the
energy, being of approximately  one order of magnitude for  $E_p =
10^{11}$ GeV. In Fig. \ref{fig:7} we present the CGC predictions for the
$x_F$-distribution and different values of energy. The shape of the
distribution becomes a little steeper as the energy increases. In
order to allow future studies of the prompt lepton fluxes
considering saturation physics we have parameterized the
$x_F$-distribution for charm production considering the following
general form
\begin{eqnarray}
& & x_F\,\frac{d\sigma(p+Air\rightarrow c+X)}{dx_F} =
A\,x_F^{\alpha}\,(1-x_F^{1.2})^n \,\,, \label{para}
\end{eqnarray}
with the quantities
\begin{eqnarray}
A  =  a_0+\left[a_1\ln\,\left(\frac{E_p}{10^8\,\mathrm{GeV}}\right)\right],\hspace{0.2cm} \alpha  =  b_0-b_1\,\ln \left(\frac{E_p}{10^4\,\mathrm{GeV}}\right),\hspace{0.2cm} n  =   n_0 + n_1\,\ln
\left(\frac{E_p}{10^4\,\mathrm{GeV}}\right).\nonumber
\end{eqnarray}
The parameters for two distinct energy ranges are presented in the
Table \ref{tab:1}.


\begin{table}[t]
\begin{center}
\begin{tabular} {||l|c|c|c|c|c|c||}
\hline \hline
 Energy Range   & $a_0$ & $a_1$ & $b_0$ & $b_1$ & $n_0$ & $n_1$ \\
\hline \hline
 $10^4<E_p<10^8$ GeV & 6804 $\mu$b & 826 & 0.05 & 0.016 & 0.075 & -0.107 \\
\hline \hline
 $10^8<E_p<10^{11}$ GeV & 5422 $\mu$b & 403 & 0.025 & 0.023 & 6.7 & -0.102 \\
\hline \hline
\end{tabular}
\end{center}
\caption{\it Parameters of the fit to the differential cross section
$x_F\,d\sigma/dx_F$ for charm production in proton-air collisions
(see text).} \label{tab:1}
\end{table}

\FIGURE{\epsfig{file=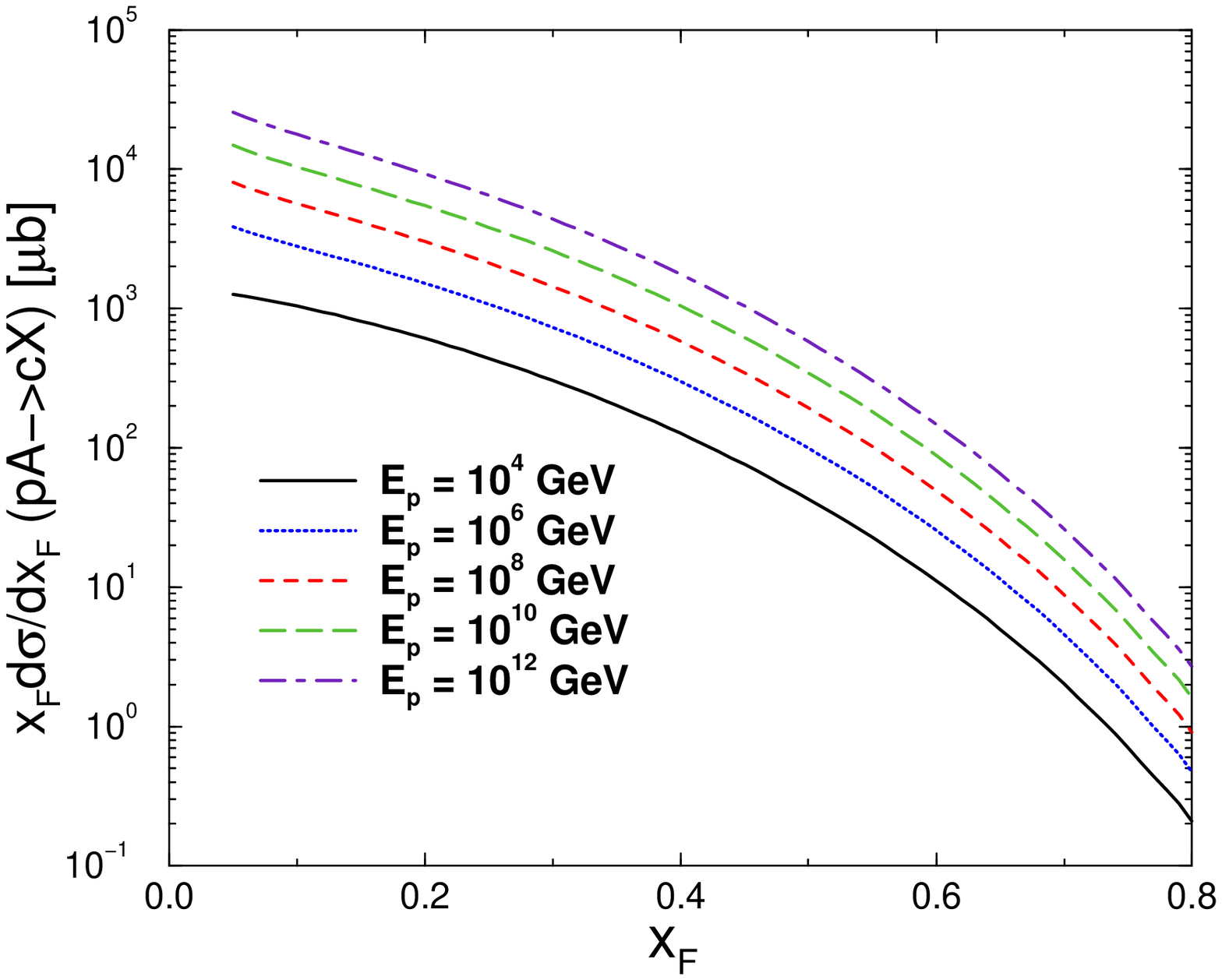,width=10cm}
        \caption[$x_F$-dependence for charm production in $pA$ interactions predicted by saturation physics.]{$x_F$-dependence for
        charm production in $pA$ interactions  predicted by saturation physics.}%
    \label{fig:7}}

Let us consider now the bottom production cross section and its
$x_F$-dependence considering the color dipole picture and saturation
physics. From the previous discussion we have that these effects are
expected to be minimal and that the cross section is dominated by
the color transparency regime. However, we present our predictions
obtained using color dipole picture and an eikonalized dipole-nuclei
cross section, in order to establish the results which comes from
this framework. As emphasized in Ref. \cite{leptonflux_mrs}, the
high energy production of $\nu_\tau$ from beauty decays is not
negligible, what motivates our study. In Fig. \ref{fig:8} we present
our predictions for the energy dependence of the total bottom
production cross section considering the CGC and color transparency
limit. As expected, we have that in the energy range considered in
this work both predictions are almost identical for proton-proton
and proton-nucleus interactions. In Fig. \ref{fig:9} the
$x_F$-distribution is shown for different proton energies. The
behavior is similar to that obtained for charm production
considering the color transparency limit, with the distribution
becoming steeper for larger energies. As for the charm case we have
parameterized this distribution considering the general form given
in Eq. (\ref{para}). The resulting parameters are presented in the
Table \ref{tab:2} for two distinct energy ranges. It important to
emphasize that similarly to the photoproduction case  the bottom
cross section increases with the energy faster than the charm one,
due to distinct physics which is dominant in the production process.
In order to illustrate this feature, we have estimated the  ratio of
the bottom and charm cross sections for proton-air collisions as a
function of proton energy for $E_p\geq 10^8$ GeV and verified  that
it is proportional to  $R_{b/c}\propto E_p^{\,0.1}$. Basically, we have that the
bottom production cross section is increasingly important and
reaches 10-13 \% of the charm cross section for ultra-high energy
cosmic rays. Consequently, the contribution for the prompt lepton
fluxes associated to the bottom becomes important for larger
energies.

\FIGURE{\epsfig{file=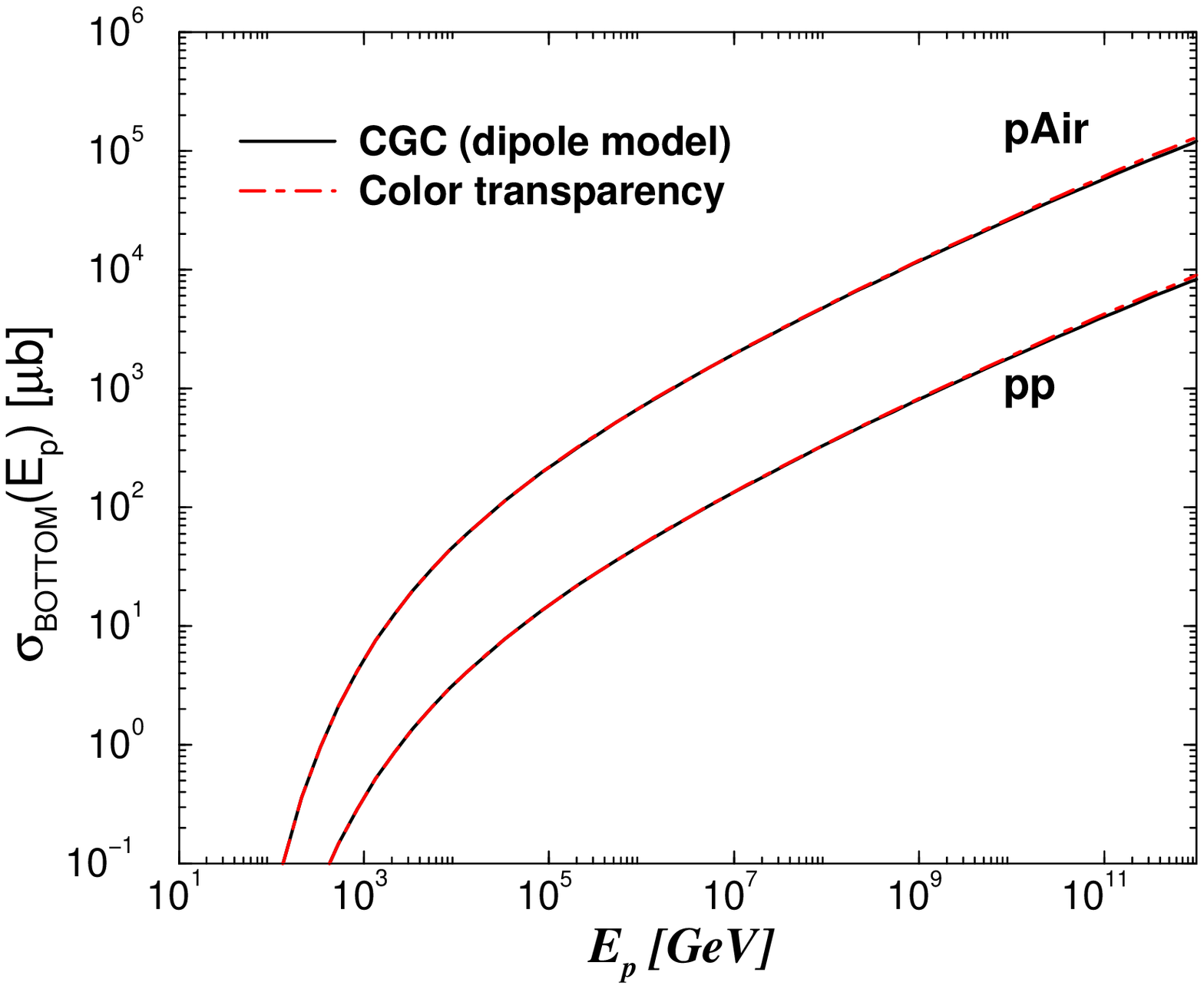,width=10cm}
        \caption[Bottom production in $pp$ and $pA$ interactions.]{Bottom production in $pp$ and $pA$ interactions.}%
    \label{fig:8}}

\FIGURE{\epsfig{file=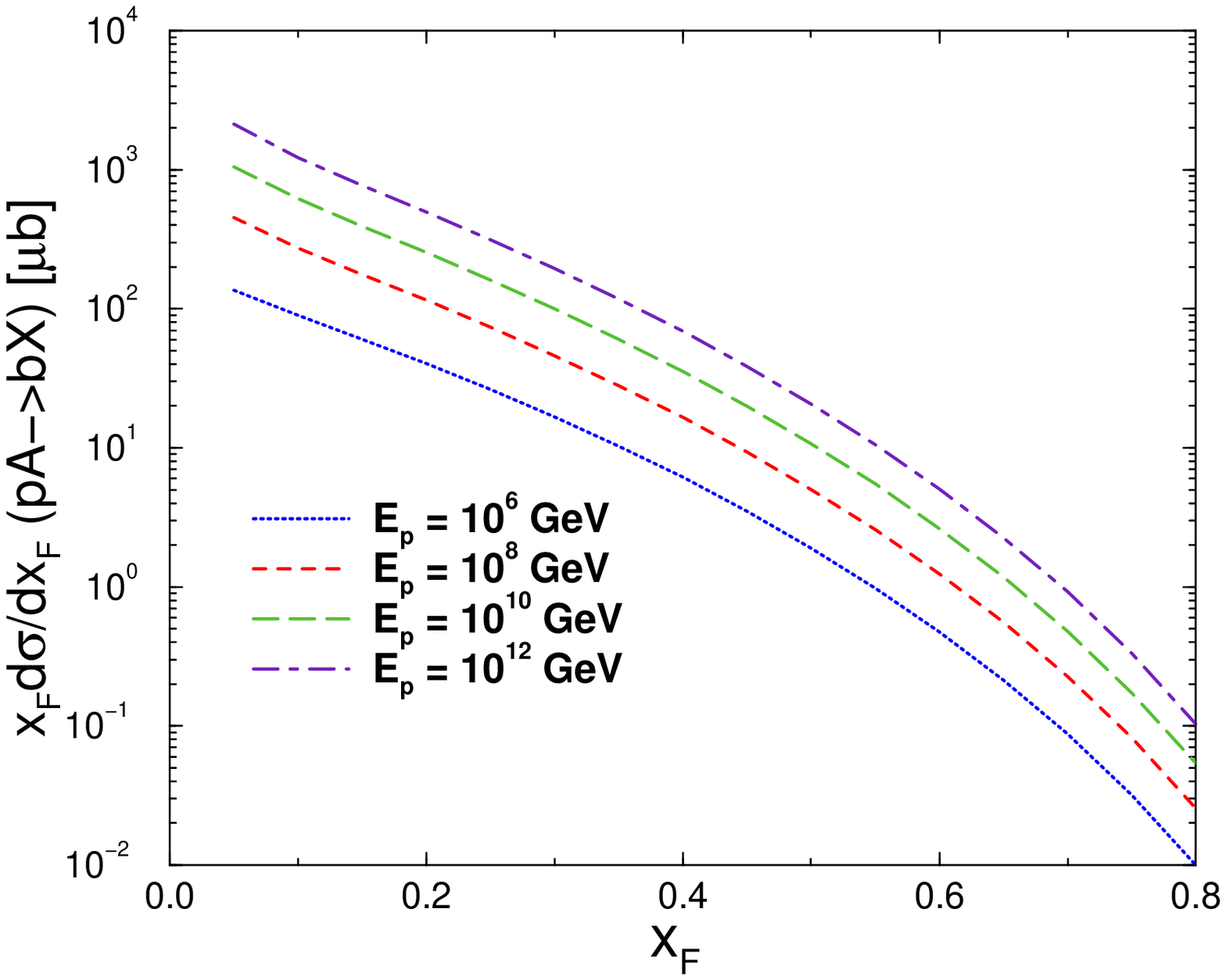,width=10cm}
        \caption[$x_F$-dependence for bottom production in $pA$ interactions.]{$x_F$-dependence for bottom production in $pA$ interactions.}%
    \label{fig:9}}


\begin{table}[t]
\begin{center}
\begin{tabular} {||l|c|c|c|c|c|c||}
\hline \hline
 Energy Range   & $a_0$ & $a_1$ & $b_0$ & $b_1$ & $n_0$ & $n_1$ \\
\hline \hline
 $10^4<E_p<10^8$ GeV &  180.23 $\mu$b & 19.79 & -0.048 & 0.035 & 7.54 & -0.091 \\
\hline \hline
 $10^8<E_p<10^{11}$ GeV & 236.13 $\mu$b & 36.20 & -0.061 & 0.023 & 9.42 & -0.18\\
\hline \hline
\end{tabular}
\end{center}
\caption{\it Parameters of the fit to the differential cross section
$x_F\,d\sigma/dx_F$ for bottom production in proton-air collisions
(see text).} \label{tab:2}
\end{table}

\section{Summary.}
\label{section5}


The determination of the prompt lepton flux is fundamental in order
to establish, for instance, the background for ultra high energy
neutrinos from cosmological sources. At high energies the lepton
fluxes are quite sensitive to the heavy quark cross section, which
implies that the choice of an appropriate theoretical framework to
estimate the cross section in this energy range is fundamental.
Previous calculations have considered perturbative QCD at
next-to-leading order, assuming the validity of the collinear
factorization and of the DGLAP dynamics in the kinematical range of
very high energies. However, current accelerator data already have
indicated the presence of new dynamical effects, associated to
saturation physics. As the contribution of these effects increases
with the energy, we can expect that it cannot be disregarded in the
description of the interaction of ultra high energy cosmic rays with
the atmosphere. Here we have estimated the heavy  quark production
in the interaction of cosmic rays in the atmosphere taking the
primary cosmic ray as a proton or a photon.  At ultra high energies
and charm production the saturation scale stays above the semihard
scale $\mu_c^2$ and the process contains sizable contribution from
the saturation regime. In particular, geometric scaling for charm
production on the scaling variable $\tau_c$ is demonstrated using
small-$x$ DESY-HERA data. Within the color dipole approach and CGC
formalism for dipole-target interaction, the parton saturation
corrections are huge, suppressing the cross section by one order of
magnitude for ultra-high energy primaries at $E_{\gamma}\approx
10^6$ GeV.
 We  also predict a factor ten of
suppression of the prompt lepton fluxes for a pure proton component
for the primary cosmic ray, increasing if the primary component
changes as appropriate for a Top-Down model. This suppression is
also present in the $x_F$-distribution, which one the main inputs to
the calculate the prompt lepton fluxes. The resulting predictions
for the total cross section at very high energies are similar to
those obtained previously by Thunman, Ingelman and Gondolo, which
have been considered a lower bound. As there is a strict relation
between the charm production and the prompt lepton fluxes, we
believe that the resulting lepton fluxes obtained using our
predictions for charm production as input of the atmospheric
particle showers routines should be similar. In other words, we
expect a suppression of the prompt lepton fluxes associated to the
saturation physics when compared with those resulting from NLO
calculations using the collinear factorization. Of course, a more
detailed analyzes is necessary in order to quantify precisely this
suppression. It is important to emphasize that these  predictions
should be considered a lower bound in the suppression, since in the
CGC formalism the breaking of the factorization for heavy quark
production is predicted for $Q_{\mathrm{sat}} \gg \mu_c$
\cite{gelis}, which implies a larger suppression.

We also estimate the bottom production considering the dipole
picture and saturation physics. In this case,  we have that the
saturation effects can be disregarded  and the color transparency
limit determines the behavior of the total cross section in the
energy range of interest in this paper. However, as the $B$-hadron
decays should contribute significantly for the flux of high energy
$\nu_\tau$ neutrinos, the quantification of this cross section using
the color dipole picture, which is expressed in terms of the
eigenstates of interaction in QCD, is useful.

We present a parameterization for the $x_F$-distribution for charm
and bottom production resulting of our calculations. It can be
useful for future calculations of the prompt lepton fluxes. The main
conclusion of our phenomenological analyzes is that the saturation
effects implies  a suppression of prompt lepton fluxes. Of course,
in a full calculation we should include the fragmentation of the
heavy quark pairs into hadrons and their subsequent semileptonic
decays, as well as, the propagation of the high energy particles
through the atmosphere. We postponed these analyzes for a future
publication.

\acknowledgments This work was  partially financed by the Brazilian
funding agencies CNPq and FAPERGS. The authors thank S. Munier and Frank   Steffen  for
helpful comments.


\begin{thebibliography}{999}

\bibitem{review_cosmic}
  P.~Bhattacharjee and G.~Sigl,
  Phys.\ Rept.\  {\bf 327}, 109 (2000); L.~Anchordoqui {\it et al.},
  Annals Phys.\  {\bf 314}, 145 (2004)



\bibitem{photonsgzk}
  G.~Gelmini, O.~Kalashev and D.~V.~Semikoz,
  arXiv:astro-ph/0506128.

\bibitem{bugaev}
  E.~V.~Bugaev, A.~Misaki, V.~A.~Naumov, T.~S.~Sinegovskaya, S.~I.~Sinegovsky and N.~Takahashi,
  Phys.\ Rev.\  D {\bf 58}, 054001 (1998)

\bibitem{review_halzen}
  T.~K.~Gaisser, F.~Halzen and T.~Stanev,
  Phys.\ Rept.\  {\bf 258}, 173 (1995)
  [Erratum-ibid.\  {\bf 271}, 355 (1996)]


\bibitem{leptonflux_tig}
  P.~Gondolo, G.~Ingelman and M.~Thunman,
  Astropart.\ Phys.\  {\bf 5}, 309 (1996).

\bibitem{leptonflux_prs}
 L.~Pasquali, M.~H.~Reno and I.~Sarcevic,
  Phys.\ Rev.\ D {\bf 59}, 034020 (1999).

 \bibitem{leptonflux_ggv}
 G.~Gelmini, P.~Gondolo and G.~Varieschi,
  Phys.\ Rev.\ D {\bf 61}, 036005 (2000).

\bibitem{leptonflux_mrs}
A.~D.~Martin, M.~G.~Ryskin and A.~M.~Stasto,
  Acta Phys.\ Polon.\ B {\bf 34}, 3273 (2003).

\bibitem{costa}
  C.~G.~S.~Costa,
  Astropart.\ Phys.\  {\bf 16}, 193 (2001)



 \bibitem{ggv_gluon}
 G.~Gelmini, P.~Gondolo and G.~Varieschi,
  Phys.\ Rev.\ D {\bf 61}, 056011 (2000).

\bibitem{dglap}
 V. N. Gribov and L.N. Lipatov, { Sov. J. Nucl. Phys} {\bf 15}, 438 (1972);
 Yu. L. Dokshitzer, { Sov. Phys. JETP} {\bf 46}, 641 (1977);
 G. Altarelli and G. Parisi, { Nucl. Phys.} {\bf B126}, 298 (1977).

\bibitem{hdqcd}
E.~Iancu and R.~Venugopalan,
arXiv:hep-ph/0303204;
 V.~P.~Goncalves and M.~V.~T.~Machado,
  Mod.\ Phys.\ Lett.\  {\bf 19}, 2525 (2004);  H.~Weigert,
  Prog.\ Part.\ Nucl.\ Phys.\  {\bf 55}, 461 (2005); J.~Jalilian-Marian and Y.~V.~Kovchegov, Prog.\ Part.\ Nucl.\ Phys.\  {\bf 56}, 104 (2006).


\bibitem{gelis}
  H.~Fujii, F.~Gelis and R.~Venugopalan,
     Phys.\ Rev.\ Lett.\  {\bf 95}, 162002 (2005);  Nucl.\ Phys.\  A {\bf 780}, 146
     (2006).

\bibitem{nik_nonlinear}
  N.~N.~Nikolaev and W.~Schafer,
  Phys.\ Rev.\  D {\bf 71}, 014023 (2005)


\bibitem{nik} N. N. Nikolaev, B. G. Zakharov,  Phys. Lett. B  {\bf 260}, 414 (1991);
{Z. Phys. C} {\bf 49}, 607 (1991).



\bibitem{npz}
N.~N.~Nikolaev, G.~Piller and B.~G.~Zakharov,
 Z.\ Phys.\ A {\bf 354}, 99 (1996);
 B.~Z.~Kopeliovich and A.~V.~Tarasov,
Nucl.\ Phys.\ A {\bf 710}, 180 (2002).



\bibitem{rauf}
  J.~Raufeisen and J.~C.~Peng,
  Phys.\ Rev.\ D {\bf 67}, 054008 (2003)










\bibitem{CGC}  I. I. Balitsky,   Nucl. Phys. {\bf  B463}, 99 (1996);
 J. Jalilian-Marian, A. Kovner and  H.
Weigert, Phys. Rev. D {\bf 59}, 014014 (1999), {\it ibid.} {\bf 59},
014015 (1999), {\it ibid.} {\bf 59}  034007 (1999); E. Iancu, A.
Leonidov and L. McLerran, Nucl.Phys.  {\bf A692} (2001) 583;
 H. Weigert, Nucl. Phys.  {\bf A703}, 823 (2002).


\bibitem{KOVCHEGOV}
Y.V. Kovchegov,  Phys. Rev. D {\bf 60},  034008 (1999).


\bibitem{IIM}
  E.~Iancu, K.~Itakura and S.~Munier,
  Phys.\ Lett.\ B {\bf 590}, 199 (2004).


\bibitem{armesto}
N. Armesto, Eur. Phys. J. C {\bf 26}, 35 (2002).

\bibitem{Shoshi:2002in}
  A.~I.~Shoshi, F.~D.~Steffen and H.~J.~Pirner,
  Nucl.\ Phys.\ A {\bf 709}, 131 (2002).

\bibitem{SGK} A. M.  Sta\'sto, K. Golec-Biernat and J. Kwieci\'nski,  Phys. Rev. Lett. {\bf 86}, 596 (2001).


\bibitem{prl}
  V.~P.~Goncalves and M.~V.~T.~Machado,
  Phys.\ Rev.\ Lett.\  {\bf 91}, 202002 (2003).


\bibitem{datahera}
S. Aid {\it et al.} (H1 collaboration), Nucl. Phys.{\bf B472}, 32
(1996); \\ C. Adloff {\it et al.} (H1 collaboration), Phys. Lett. B
{\bf 467}, 156 (1999).


\bibitem{fixedtarget}

M.S. Atiya {\it et al.}, Phys. Rev. Lett. {\bf 43}, 414 (1979);\\
D. Aston {\it et al.} (WA4 collaboration), Phys. Lett. B{\bf 94}, 113 (1980);\\
J.J. Aubert {\it et al.} (EMC collaboration), Nucl. Phys. {\bf B213}, 31 (1983);\\
K. Abe {\it et al.} (SHFP collaboration), Phys. Rev. Lett. {\bf 51}, 156 (1983);\\
K. Abe {\it et al.} (SHFP collaboration), Phys. Rev. D {\bf 33}, 1 (1986);\\
M.I. Adamovich, Phys. Lett. B {\bf 187}, 437 (1987);\\
J.C. Anjos {\it et al.} (The Tagged Photon Spectrometer
collaboration),
Phys. Rev. Lett. {\bf 65}, 2503 (1990);\\
J.J.~Aubert {\it et al.}  [European Muon Collaboration], Phys.\
Lett.\ B {\bf 106}, 419 (1981).

\bibitem{MUNIER}
S.~Munier, Phys.\ Rev.\ D {\bf 66}, 114012 (2002).


\bibitem{ASW}
  N.~Armesto, C.~A.~Salgado and U.~A.~Wiedemann,
  Phys.\ Rev.\ Lett.\  {\bf 94}, 022002 (2005).



\bibitem{mrst2004nlo} A. D. Martin {\it et al.}, Phys. Lett. B {\bf 604}, 61 (2004).

\bibitem{accdata}
K.~Kodama {\it et al.}, Phys.\ Lett.\ B {\bf 263}, 573 (1991);
R.~Ammar {\it et al.}, Phys.\ Rev.\ Lett.\  {\bf 61}, 2185 (1988);
M.~Aguilar-Benitez {\it et al.}, Z.\ Phys.\ C {\bf 40}, 321 (1988);
G.~A.~Alves {\it et al.}, Phys.\ Rev.\ Lett.\  {\bf 77}, 2388 (1996)
[Erratum-ibid.\  {\bf 81}, 1537 (1998)]; S.~Barlag {\it et al.}, Z.\
Phys.\ C {\bf 39}, 451 (1988); K.~Adcox {\it et al.}, Phys.\ Rev.\
Lett.\  {\bf 88}, 192303 (2002); O. Botner {\it et al.}, Phys.\
Lett.\ B {\bf 236}, 488 (1990).





\end{thebibliography}
\end{document}